\documentclass[
a4paper,amsfonts,amssymb,amsmath,reprint,showkeys,oneside,
pdftex,colorlinks=true,linkcolor=blue,citecolor=blue,
urlcolor=blue,superscriptaddress,nofootinbib
]{revtex4-1}
\newcommand{\mathd}{\mathrm{d}}
\newcommand{\vol}{\operatorname{vol}}

\newcommand{\BV}{\mathbb V}
\newcommand{\BC}{\mathbb C}

\newcommand{\BZ}{\mathbb Z}

\newcommand{\dd}{\mathd}
\newcommand{\ii}{\mathrm{i}}
\allowdisplaybreaks[1]

\usepackage[utf8]{inputenc}
\usepackage[colorinlistoftodos, color=green!40, prependcaption]{todonotes}
\usepackage{amsmath}
\usepackage{slashed}
\usepackage[english]{babel}
\usepackage{graphicx}
\usepackage[pdftex, pdftitle={Article}, pdfauthor={Author}]{hyperref}
\usepackage{geometry}
\begin{document}

\title{Probe branes from equivariant volumes}

\author{Kiril Hristov}
\affiliation{School of Natural Sciences, Institute for Advanced Study, 1 Einstein Drive, Princeton, NJ 08540, USA}
\affiliation{Faculty of Physics, Sofia University, J. Bourchier Blvd. 5, 1164 Sofia, Bulgaria}
\affiliation{INRNE, Bulgarian Academy of Sciences, Tsarigradsko Chaussee 72, 1784 Sofia, Bulgaria}

\author{Pedro Vicente Marto}
\affiliation{\mbox{Institute for Theoretical Physics,
Utrecht University, Princetonplein 5, 3584 CC Utrecht, The Netherlands}}

\begin{abstract}
\noindent Classical Euclidean probe actions control non-perturbative
weights and extended observables in holography. We present
a geometric prescription for M2-branes in arbitrary BPS
backgrounds and D3-branes in the five-form sector of type IIB.
The framework accommodates compact and non-compact supersymmetric
worldvolumes in both internal and spacetime directions,
expressing their actions as relative equivariant pairings.
Under stated extension and boundary assumptions, equivariant
volumes evaluate these pairings without explicit background
metrics and provide a regularization for non-compact probes.
Examples in $\mathrm{AdS}_4\times SE_7$ backgrounds
dual to ABJM, ADHM and flavored quiver theories
reproduce and refine worldsheet-type instanton exponents
and finite Wilson-loop actions.
We also recover M2 giant-graviton and surface-defect
actions in $\mathrm{AdS}_7\times S^4$.
D3 probes in $SE_5$ compactifications dual to
$\mathcal{N}=4$ super Yang--Mills and Klebanov--Witten
theory provide tests involving giant gravitons,
black hole probes and surface defects.
\end{abstract}

\maketitle

\section{Introduction}
\label{sec:intro}

Euclidean probe branes describe non-perturbative sectors of
holographic partition functions and indices, as well as extended
operators when their worldvolumes intersect the conformal boundary.
Their classical actions determine exponential weights, while
fluctuations and zero modes determine the prefactors.
Direct string and M2-brane calculations
\cite{Gautason:2023igo,Beccaria:2023ujc,Beccaria:2023sph}
and calibration and localization methods
\cite{Kurlyand:2026yke,Kurlyand:2026gxi,BenettiGenolini:2026hmz}
establish such contributions for various probe families.
Non-compact examples include Wilson loops and surface
defects with known renormalized actions
\cite{Farquet:2013cwa,Farquet:2014bda,Gautason:2025bft,Chen:2023lzq}. Our goal is to organize these classical probe actions using
the equivariant data of the background.

Equivariant volumes connect Sasaki--Einstein geometry to
Calabi--Yau cones \cite{Martelli:2005tp,Martelli:2006yb}
and provide regularized integrals on non-compact spaces
\cite{Martelli:2023oqk,Cassia:2025aus,Cassia:2025jkr,Hristov:2026zjh}.
Related contact volume formulas determine holographic
free energies \cite{Gabella:2011sg}.
Building on anomaly inflow
\cite{Freed:1998tg,Harvey:1998bx,Bah:2019rgq,Bah:2020jas},
the construction of \cite{Hristov:2026uxs} organizes
background on-shell actions through an auxiliary geometry $Y_{12}$ and an
equivariant flux class $E_4$ or $E_6$.
The torus weights satisfy the Calabi--Yau constraint
of the auxiliary construction.
We denote the equivariant-volume prescription by
$\int^{\mathbb V}$, reviewed in Appendix \ref{app:equivariant}, with insertions evaluated by derivatives of the volume generating
function. The background action representatives are~\footnote{In
IIB we retain only five-form flux, with constant axion
and dilaton, and restrict to mesonic background data.
We assume that $Y_{12}=X_6\times Z_6$ models compatible
fillings of the external and internal five-dimensional
spaces, with $E_6$ here a magnetic flux representative.}
\begin{align}
\label{eq:backgroundfunctionsM}
 I_{\rm M}&=2\pi\ii\int_{Y_{12}}^{\mathbb V}
       \left(\frac16E_4^3+E_4I_8\right)\,,\\
 I_{\rm IIB}&=2\pi\ii\int_{Y_{12}}^{\mathbb V}
       \frac12E_6^2\,.
 \label{eq:backgroundfunctionsIIB}
\end{align}
Here $I_8=(p_1^2-4p_2)/192$ uses equivariant
Pontryagin classes \cite{Freed:1998tg}.

We connect this prescription to relative equivariant
localization of physical actions
\cite{BenettiGenolini:2026cdw,Couzens:2026xmi}, using
relative classes, their products and equivariant Thom forms
\cite{ParadanVergne2007,Fujisawa2018}.
For compact fillings satisfying the localization hypotheses,
the complete relative pairing is evaluated by the
Atiyah--Bott--Berline--Vergne formula, denoted $\int^{\rm eq}$.
We represent the required fillings by auxiliary geometric
models. On non-compact models, $\int^{\mathbb V}$ defines
a meromorphic pushforward, finite at generic weights.
Its identification with the finite physical action,
including the required boundary terms, is an assumption
tested by the examples below.

Keeping the background data fixed, we propose the
actions of individual probes:
\begin{align}\label{eq:M2generalpushforward}
 I_{\rm M2}&=2\pi\ii\int_{D_4}^{\mathbb V}\iota_D^*E_4\,, \\ 
 I_{\rm D3}&=2\pi\ii\int_{D_6}^{\mathbb V}\iota_D^*E_6\,.
\end{align}
Here $\iota_D:D\hookrightarrow Y_{12}$ is the equivariant
inclusion, with subscripts indicating real dimensions.
The cycles model fillings of calibrated worldvolumes,
with compatible orientations and potential data.
For M2-branes, relative descent of the flux polyform
uses a four-dimensional filling of the worldvolume.
The D3 construction uses a product of two relative
fillings of odd-dimensional worldvolume factors,
with the physical worldvolume as a corner of the six-dimensional filling.

Compatible equivariant parameters encode supersymmetric
deformations, including squashing, complexified masses
and chemical potentials
\cite{Cassia:2025aus,Cassia:2025jkr,Hristov:2026zjh,Hristov:2026uxs}.
Their physical interpretation presupposes a supersymmetric
background and a probe embedding preserving the chosen
supercharge.
Our prescription evaluates the resulting actions without
solving the general existence or classification problem
for backgrounds and embeddings.
We therefore distinguish checks based on known calibrated
embeddings and geometric constructions from refinements
conditional on a compatible supersymmetric realization.

\textbf{Note added:}
While this manuscript was being finalized,
\cite{BenettiGenolini:2026gjn} appeared, whose Section~2
overlaps with our derivation of the probe M2-brane action
using relative equivariant localization in
Section~\ref{sec:M2action}.~\footnote{An earlier version
of this manuscript, containing this derivation along
with the rest of our main results, was timestamped
on 12~November~2026. Documentation of the timestamp
accompanies our arXiv submission. This earlier draft
is available from the authors upon request.}

\section{M2-branes}
\label{sec:M2}

\subsection{M-theory action}

We begin with the relation between the physical action and
the auxiliary class of \cite{Hristov:2026uxs}. Using the trace of Einstein's equation,
the two-derivative Euclidean M-theory bosonic action
takes the bulk on-shell form \cite{BenettiGenolini:2026cdw}
\begin{equation}
 I_{\rm M}^{(2\partial)}
 =\frac1{16\pi G_{11}}\int_{M_{11}}
 \left(\mathcal L_{\rm GI}+\mathcal L_{\rm CS}\right)\,,
 \label{eq:Mtheoryaction}
\end{equation}
where $\mathcal L_{\rm GI}=G_4\wedge *G_4/3$ and
$\mathcal L_{\rm CS}=\ii C_3\wedge G_4^2/6$.
Locally $G_4=\dd C_3$; globally, the Chern--Simons
coupling requires compatible flux and extension data
\cite{Witten:1996md,Freed:1998tg}.

Let $\varepsilon$ be a background Killing spinor.
We use analytically continued charge-conjugation bilinears,
with $\bar\varepsilon=\varepsilon^T\mathcal C$ and
$\mathcal C$ the charge-conjugation matrix.
Writing
$\Gamma_{(r)}=\frac1{r!}\Gamma_{M_1\cdots M_r}
\dd x^{M_1}\wedge\cdots\wedge\dd x^{M_r}$,
the associated Killing vector and form bilinears are
\cite{BenettiGenolini:2026cdw}
\begin{equation}
 K^M=\bar\varepsilon\Gamma^M\varepsilon\,,\qquad
 \Omega_2=\bar\varepsilon\Gamma_{(2)}\varepsilon\,,\qquad
 \Sigma_5=\bar\varepsilon\Gamma_{(5)}\varepsilon\,.
 \label{eq:Mtheorybilinears}
\end{equation}
With $\dd_K=\dd-\iota_K$ acting on invariant polyforms,
supersymmetry and the flux Bianchi identity and equation of motion imply
\cite{Gauntlett:2002fz,BenettiGenolini:2026cdw}
\begin{equation}
 \begin{aligned}
 \Phi_G&=G_4+\Omega_2\,,\qquad \dd_K\Phi_G=0\,,\\
 \Phi_{\rm GI}&=\frac13\Phi_G\wedge(*G_4+\ii\Sigma_5)\,,\\
 \dd_K\Phi_{\rm GI}&=-\frac{\ii}{6}\Phi_G^3\,.
 \end{aligned}
 \label{eq:Mtheorypolyforms}
\end{equation}
Their top components are $[\Phi_G]_4=G_4$ and
$[\Phi_{\rm GI}]_{11}=\mathcal L_{\rm GI}$.
Importantly, the last identity relates the gauge-invariant action
to the equivariant completion of the Chern--Simons anomaly
form, as required for the relative pairing in
\eqref{eq:relativepair}.

Suppose $\Phi_G$ admits a compatible equivariantly closed
extension $\Phi_G^Y$ to a filling $Y_{12}$ of $M_{11}$.
The pair
\begin{equation}
 \alpha=\frac{\ii(\Phi_G^Y)^3}{96\pi G_{11}}\,,\qquad
 \beta=\frac{\Phi_{\rm GI}}{16\pi G_{11}}
 \label{eq:Mbulkrelativepair}
\end{equation}
then gives $I_{\rm M}^{(2\partial)}=J[\alpha,\beta]$ (see notation defined in Appendix \ref{app:equivariant}).
The extension includes the chosen potential data and may
have a zero-form component absent from $\Phi_G$.
With $16\pi G_{11}=(2\pi)^8\ell_p^9$, the identification
with the auxiliary flux class of \cite{Hristov:2026uxs} is
\begin{equation}
 E_4\simeq\frac{\Phi_G^Y}{(2\pi\ell_p)^3}\,,
 \label{eq:Mfluxdictionary}
\end{equation}
where $\simeq$ denotes equality in equivariant cohomology,
rather than equality of representatives. Compatible
relative boundary data are also required when writing \eqref{eq:Mfluxdictionary}.

For compact $Y_{12}$ with closed boundary
$M_{11}=\partial Y_{12}$ and $K$ nowhere zero there,
relative localization evaluates this pairing with $\int^\mathrm{eq}$. With the volume prescription
for non-compact auxiliary models, the same expression is
\begin{equation}
 I_{\rm M}^{(2\partial)}
 =2\pi\ii\int_{Y_{12}}^{\mathbb  V}\frac{E_4^3}{6}\,.
 \label{eq:Mbulkaction}
\end{equation}
The equality includes the boundary prescription stated
in the introduction. The anomaly coupling supplies the
$E_4I_8$ term in
\eqref{eq:backgroundfunctionsM} through the corresponding
higher-derivative relative construction
\cite{BenettiGenolini:2026cdw,BenettiGenolini:2026kjy}.~\footnote{The anomaly form and its equivariant completion
are known, while the descent identity for the full supersymmetric
higher-derivative gauge-invariant action remains conjectural.}

\subsection{Probe M2-brane action}
\label{sec:M2action}

Consider an M2-brane wrapping a compact oriented
three-manifold $\Sigma_3\subset M_{11}$ without boundary.
Let $P$ denote pullback to $\Sigma_3$ and
$\vol_{\Sigma_3}$ the volume form of its induced metric.
The bosonic part of the kappa-symmetric supermembrane
action, in our Euclidean conventions, is
\cite{Bergshoeff:1987cm,Bergshoeff:1987qx}
\begin{equation}
 I_{\rm M2}
 =T_2\int_{\Sigma_3}
   \bigl(\vol_{\Sigma_3}+\ii P[C_3]\bigr)\,,
 \label{eq:M2action}
\end{equation}
where $T_2=[(2\pi)^2\ell_p^3]^{-1}$.
The M2 has no independent worldvolume gauge field.

Preserving $\varepsilon$ imposes the Euclidean
$\kappa$-symmetry condition, which implies the calibration
identity
\cite{Becker:1995kb,Gutowski:1999tu,Hackett-Jones:2003hho}:~\footnote{\label{ft:M2}In an oriented orthonormal tangent frame,
$\Gamma_{\Sigma_3}
=\frac{1}{3!}\epsilon^{abc}\gamma_{abc}
=\gamma_1\gamma_2\gamma_3$, so
$\gamma_1\Gamma_{\Sigma_3}=\gamma_{23}$.
The projector therefore gives
$K^1=\ii\bar\varepsilon\gamma_{23}\varepsilon
=\ii(P[\Omega_2])_{23}$, and cyclically.
These are the components of the second identity.
The projector and Wess--Zumino coupling use the
same charge and orientation convention.}
\begin{equation}
 \ii\Gamma_{\Sigma_3}\varepsilon=\varepsilon
 \quad \Rightarrow \quad \iota_K\vol_{\Sigma_3}=\ii P[\Omega_2]\,.
 \label{eq:M2calibration}
\end{equation}
The $\kappa$-symmetry condition also implies that the
Killing vector $K$ associated with the preserved spinor
is tangent to $\Sigma_3$
\cite[Sec.~5.2]{Figueroa-OFarrill:2004qhu}.
We additionally require $K$ to be nowhere zero there.
Invariance of the worldvolume under this vector is
necessary but does not by itself ensure the
$\kappa$-symmetry condition.

Now assume that the physical embedding admits a compatible
equivariant filling $W_4\subset Y_{12}$, with
$\partial W_4=\Sigma_3$ and boundary inclusion $j$.
Write $\Phi_G^W=\Phi_G^Y|_{W_4}$.
The extension must reproduce both the boundary polyform,
$j^*\Phi_G^W=P[\Phi_G]$, and the potential coupling:
\begin{equation}
 \int_{W_4}[\Phi_G^W]_4
 =\int_{\Sigma_3}P[C_3].
 \label{eq:M2potentialextension}
\end{equation}
This equality is understood for fixed global potential
data and a chosen branch of the Wess--Zumino
action.~\footnote{Patchwise potentials require transition
data compatible with flux quantization
\cite{Witten:1996md,Freed:1998tg}.
Flat potential holonomies are not determined by the
field strength alone.}
The relative equivariant pair is
\begin{equation}
 \alpha=\ii T_2\Phi_G^W\,,\qquad
 \beta=T_2\vol_{\Sigma_3}\,.
 \label{eq:M2relativepair}
\end{equation}
From \eqref{eq:Mtheorypolyforms}, equivariant closure gives $\dd_K\alpha=0$, while
\eqref{eq:M2calibration} gives
$\dd_K\beta=-\ii T_2P[\Omega_2]=-j^*\alpha$.
The four-form pulls back to zero on $\Sigma_3$, while the scalar component vanishes there
by matching with the physical boundary data.
We conclude that $I_{\rm M2}=J[\alpha,\beta]$, thus realizing the probe action in terms of a relative pairing.

Let $D_4\subset Y_{12}$ model the filling $W_4$
in the volume prescription, with compatible
orientation and relative potential data.
Using \eqref{eq:Mfluxdictionary},
$T_2(2\pi\ell_p)^3=2\pi$, and the projection
formula \eqref{eq:projectionformula}, we obtain
\begin{equation}
 \begin{aligned}
 I_{\rm M2}
 =2\pi\ii\int_{D_4}^{\mathbb V}\iota_D^*E_4=2\pi\ii\int_{Y_{12}}^{\mathbb V}E_4[D_4]_T\,,
 \end{aligned}
 \label{eq:M2ambientpairing}
\end{equation}
where $[D_4]_T=(\iota_D)_*1$.
Thus one may either restrict the fixed background
class to the probe cycle or insert the cycle class
in the ambient pushforward.

For non-compact physical worldvolumes, this
identification additionally requires the volume
prescription to reproduce contributions from boundary
terms regularizing the observables. The Wilson loop
examples below provide explicit checks that this is indeed the case.

\subsection{Examples}
\label{sec:M2examples}

We evaluate the probe pairing using the background class
and cycle volumes, as in \eqref{eq:M2ambientpairing}. Fixed-$N$ expressions refer to the leading order
background saddle, with torus weights and discrete data
held fixed. Following Appendix \ref{app:equivariant}, characteristic classes and invariant subcycles are inserted
by derivatives of the equivariant volume $\BV_X(\lambda,\epsilon)$, keeping the
background class fixed. We denote
$C_X(\epsilon)=\BV_X(0,\epsilon)$, with
$C_{\BC^d}=1/\prod_i\epsilon_i$.

\subsubsection{Worldsheet instantons in
$\mathrm{AdS}_4\times SE_7$}
\label{sec:M2AdS4}

We consider Euclidean AdS$_4$ with squashed three-sphere boundary. For a transverse Calabi--Yau cone $X_8=C(SE_7)$,
treated by equivariant resolution when necessary,
the auxiliary background is \cite{Hristov:2026uxs}
\begin{equation}
Y_{12}=\BC^2_{\omega_1,\omega_2}\times X_8\,,\qquad
 E_4=\frac{\mu}{2\pi\ii}e_T(T\BC^2)\,.
 \label{eq:M2AdS4background}
\end{equation}
Here $\mu$ is the chemical potential conjugate to
the M2-brane charge $N$; the distinction between fixed
potential and fixed charge is discussed in
\cite{Gautason:2025plx}.~\footnote{$N$ denotes the background M2-, M5- or D3-brane
charge, as appropriate, and $\mu$ its conjugate chemical
potential where used.} The actual
auxiliary weights are $(\omega_1,\omega_2;-\epsilon)$:
the internal holomorphic four-form has conventional weight
$\omega_1+\omega_2$, so the total signed canonical
weight vanishes by supersymmetry, $\sum_i\epsilon_i=\omega_1+\omega_2$ \cite{Hristov:2026tde,Hristov:2026uxs}.
We write cycle volumes at the conventional internal weights,
using $C_D(-\epsilon)=(-1)^{\dim_{\BC}D}C_D(\epsilon)$
when evaluating the signed pushforward.
For the squashed three-sphere we take
$\omega_1=b$, $\omega_2=b^{-1}$, with $b>0$
\cite{Cassia:2025jkr}. Since $\omega_1\omega_2=1$, the two-derivative
background action at fixed $\mu$ is
\begin{equation}
 I_{\rm M}^{(2\partial)}(\mu;\omega,\epsilon)
 =\frac{2\pi\ii}{6}\int_{Y_{12}}^{\mathbb V}E_4^3
 =-\frac{\mu^3}{24\pi^2}C_X(\epsilon).
\label{eq:M2AdS4backgroundaction}
\end{equation}

We consider internal probes wrapping Sasakian
three-cycles $\Sigma_3\subset SE_7$ whose cones
$D_4=C(\Sigma_3)\subset X_8$ are invariant holomorphic
surfaces \cite{Kurlyand:2026yke,Kurlyand:2026gxi}.~\footnote{These are candidate classical actions. The corresponding one-loop prefactors may vanish due to unsaturated fermionic zero modes or cancellations, while bosonic zero modes require
integration over collective coordinates. We keep in mind that these effects are not captured by the cycle volumes
\cite{Gautason:2023igo,Kurlyand:2026gxi,BenettiGenolini:2026gjn}.}
The radially truncated cone models the filling $W_4$,
while the full cone at the external $\BC^2$ fixed point
defines the auxiliary cycle.
The external Euler class restricts to
$\omega_1\omega_2=1$ times the unit class on $D_4$.
Equations~\eqref{eq:M2generalpushforward} and
\eqref{eq:cyclevolume} therefore give
\begin{equation}
 I_{\rm M2}(\mu;\omega,\epsilon) = \mu\, \omega_1 \omega_2\, C_D(\epsilon) 
 =\mu\, C_D(\epsilon)\,.
 \label{eq:M2AdS4action}
\end{equation}
The non-zero flux insertion comes from the equivariant
completion, since the ordinary Freund--Rubin flux
$G_4$ has no internal component.
The volume $C_D$ is computed by inserting the cycle
class of $D_4\subset X_8$ in $\BV_X$, and satisfies
$C_D(-\epsilon)=C_D(\epsilon)$.

The cubic background term fixes the leading saddle
of $I_{\rm M}^{(2\partial)}(\mu)+\mu N$, giving
\begin{equation}
 \begin{aligned}
 \mu_*&=\frac{2\pi\sqrt{2N}}
    {\sqrt{C_X(\epsilon)}}\,,\\
 I_{\rm M2}(N;\epsilon)
 &=2\pi\sqrt{2N}\,
   \frac{C_D(\epsilon)}{\sqrt{C_X(\epsilon)}}\,.
 \end{aligned}
 \label{eq:M2fixedN}
\end{equation}
The probe action scales as $N^{1/2}$, compared with
the $N^{3/2}$ scaling of the background action.
Since $C_D$ and $C_X$ are homogeneous of degrees
$-2$ and $-4$, respectively, their ratio in
\eqref{eq:M2fixedN} is of degree $0$.
Pure boundary squashing at fixed internal Reeb direction
rescales these weights uniformly, leaving the leading
fixed-$N$ probe action unchanged.
At fixed $\mu$, the probe action instead scales as
$(b+b^{-1})^{-2}$.

For ABJM \cite{Aharony:2008ug},
$X_8=\BC^4/\BZ_k$, with quotient charges
$(1,1,-1,-1)$. The coordinate surface $D_{ij}$ spanned
by two planes $z_i,z_j$ has
\begin{equation}
C_{\BC^4/\BZ_k}=\frac1{k\prod_{l=1}^4\epsilon_l},\qquad
 C_{D_{ij}}=\frac1{k\epsilon_i\epsilon_j}.
 \label{eq:ABJMrefined}
\end{equation}
At $b=1$ and $\epsilon_i=1/2$, this gives
$I_{\rm M2}=4\mu_*/k=2\pi\sqrt{2N/k}$, matching
the wrapped $S^3/\BZ_k$ action
\cite[Eq.~(4.12)]{Beccaria:2023ujc}.
The corresponding worldsheet sector is known from
the matrix model
\cite{Drukker:2010nc,Marino:2011eh,Hatsuda:2012dt}
and direct string and M2 calculations
\cite{Cagnazzo:2009zh,Gautason:2023igo,
Kurlyand:2026yke,Kurlyand:2026gxi,Tseytlin:2026dhy}.

The comparison extends beyond equal weights.
At $b=1$ and
$\epsilon_1+\epsilon_3=\epsilon_2+\epsilon_4=1$,
the actions for $D_{12},D_{14},D_{23},D_{34}$
reproduce the four mass-deformed worldsheet exponents
conjectured in \cite{Nosaka:2015iiw} and tested
numerically in \cite[Eq.~(5.7)]{Nosaka:2024gle}.~\footnote{The dictionary is
$\epsilon_{1,3}=(1\pm\ii m_1/\pi)/2$ and
$\epsilon_{2,4}=(1\pm\ii m_2/\pi)/2$.
Choosing one weight from each pair gives
$I_{\rm M2}=4\mu/
[k(1\pm\ii m_1/\pi)(1\pm\ii m_2/\pi)]$,
with independent signs.
This matches the exponents without deriving the
selection of contributing cycles.}
For pure squashing, $\epsilon_i=(b+b^{-1})/4$, $\forall i$,
so $I_{\rm M2}(\mu)=16\mu/[k(b+b^{-1})^2]$.
Its fixed-$N$ value remains $2\pi\sqrt{2N/k}$,
as in the arbitrary squashing probe calculation
\cite[Eq.~(24)]{Gautason:2025per}, whose one-loop correction
depends on $b$.
At $b^2=3$ and $k=1$, the fixed-$\mu$ action becomes
$3\mu$, matching the $e^{-3\mu}$ sector of the exact
ellipsoid result \cite[Eq.~(3.22)]{Hatsuda:2016uqa}.

For ADHM with flavor number $N_f$, we use, we use the toric resolution
$X_8=\BC^2\times A_{N_f-1}$ of
$\BC^2\times(\BC^2/\BZ_{N_f})$, where the quotient
acts with charges $(1,-1)$ on the last two coordinates.
The equivariant volume factorizes as
$\BV_X=\BV_{\BC^2}\BV_{A_{N_f-1}}$.
The ALE volume is obtained by summing its $N_f$
fixed-point contributions
\cite{Moore:1997dj,DelZotto:2021gzy}, giving
$C_{A_{N_f-1}}=1/(N_f\epsilon_3\epsilon_4)$.
\footnote{The tangent weights at fixed point
$r=0,\ldots,N_f-1$ are
$(N_f-r)\epsilon_3-r\epsilon_4$ and
$-(N_f-r-1)\epsilon_3+(r+1)\epsilon_4$.
The sum of their inverse products telescopes
to the stated result; see
\cite[Eqs.~(A.22)--(A.24)]{DelZotto:2021gzy}.} Consequently, the ambient volume and the volumes
of the four mixed coordinate surfaces are
\begin{equation}
 C_{\BC^2\times A_{N_f-1}}=\frac1{N_f\prod_{l=1}^4\epsilon_l}\,,\qquad
 C_{D_{ij}}=\frac1{N_f\epsilon_i\epsilon_j}\,,
 \label{eq:ADHMrefined}
\end{equation}
with $i=1,2$ and $j=3,4$.
The two unmixed surfaces give
\begin{equation}
C_{D_{12}}=\frac1{\epsilon_1\epsilon_2},\qquad
 C_{D_{34}}=\frac1{N_f\epsilon_3\epsilon_4}.
 \label{eq:ADHMunmixed}
\end{equation}
Here $D_{12}$ denotes the reduced $\BC^2$ surface
arising from the plane fixed pointwise by the orbifold
group. On the resolution, we represent it by $\BC^2$
times one ALE fixed point, with no orbifold stabilizer
factor. The surface $D_{34}$ is the ALE factor.
At $b=1$ and $\epsilon_i=1/2$, the mixed surfaces
and $D_{34}$ give
$I_{\rm M2}=4\mu_*/N_f=2\pi\sqrt{2N/N_f}$,
whereas $D_{12}$ gives
$I_{\rm M2}=4\mu_*=2\pi\sqrt{2N_fN}$.
For the mixed surfaces, this matches the direct
M2-brane action \cite[Eq.~(3.52)]{Kurlyand:2026gxi}
and the worldsheet exponent of the $N_f$ matrix
model \cite[Eq.~(3.22)]{Hatsuda:2014vsa},
see also \cite{Grassi:2014vwa}.
For the unmixed surfaces, a separate fluctuation
analysis is needed to determine whether these
classical configurations contribute.

For $X_8 = C(Q^{1,1,1})$, the fourfold is the affine cone
over the Segre embedding of $(\mathbb P^1)^3$, while 
its link is an $S^1$ bundle over $S^2\times S^2\times S^2$.
The corresponding flavored quiver theories were constructed
in \cite{Benini:2009qs,Jafferis:2009th}.
Following \cite{Hristov:2026zjh}, assign homogeneous-coordinate
weights $(\epsilon_1,\epsilon_2)$,
$(\epsilon_3,\epsilon_4)$ and
$(\epsilon_5,\epsilon_6)$ to the three factors, with
$\sum_{i=1}^6\epsilon_i=b+b^{-1}$.
The ambient volume follows from
\cite[Eq.~(3.69)]{Hristov:2026zjh}:
\begin{equation}
 \begin{split}
 C_{C(Q^{1,1,1})}(\epsilon)
 =&\frac1{(\epsilon_1-\epsilon_2)
             (\epsilon_3-\epsilon_4)}\\
 &\times\sum_{i=1}^2\sum_{j=3}^4(-1)^{i+j}
\prod_{l=5}^6\frac1{\epsilon_i+\epsilon_j+\epsilon_l}.
 \end{split}
 \label{eq:Q111volume}
\end{equation}
The apparent poles at equal weights within either
of the first two pairs cancel in the sum.
Fixing coordinate points in the first and third
projective factors and varying the second gives
surfaces $D_{i\ell}\simeq\BC^2$, with $i=1,2$ and
$\ell=5,6$, and volumes
\begin{equation}
 C_{D_{i\ell}}(\epsilon)
 =\frac1{(\epsilon_i+\epsilon_3+\epsilon_\ell)
          (\epsilon_i+\epsilon_4+\epsilon_\ell)}.
 \label{eq:Q111refined}
\end{equation}
Permuting the three pairs gives the other coordinate
surfaces. At $b=1$ and
$\epsilon_1=\cdots=\epsilon_6=1/3$, the smooth limit
gives $C_{C(Q^{1,1,1})}=6$ and $C_{D_{i\ell}}=1$.
Hence $I_{\rm M2}=\mu_*=2\pi\sqrt{N/3}$,
reproducing the $N_f=1$ classical action for the
monopole cycles in \cite[Eq.~(3.29)]{Kurlyand:2026gxi}.

The same volume calculation can be applied to the
$\mathcal N=4$ $(q,p)_k$ necklace quivers, with cones
$(\BC^2/\BZ_q\times\BC^2/\BZ_p)/\BZ_k$, once the
effective quotient action and cycle multiplicities are specified.
Their known worldsheet sectors and deformations provide
further comparisons
\cite{Moriyama:2014nca,Nosaka:2015iiw,Kurlyand:2026yke}. 
Beyond the explicitly checked loci, a physical interpretation
requires a supersymmetric background and embedding realizing
the chosen weights. The same prescription applies to arbitrary toric
$SE_7$, using the equivariant volumes of $C(SE_7)$
and the holomorphic surfaces whose links support
supersymmetric M2 probes. As opposed to these worldsheet instantons, the interpretation of membrane instantons in these geometries
remains less understood, but their exponents relate to quantum equivariant volumes
\cite{Hristov:2026nhx}.

\subsubsection{Wilson loops in $\mathrm{AdS}_4\times SE_7$}
\label{sec:M2Wilson}

We use the background data
\eqref{eq:M2AdS4background} with
$\omega_1=b$, $\omega_2=b^{-1}$ and $b>0$.
An M2 wrapping a supersymmetric external disk and
a closed internal orbit describes a Wilson loop,
with leading expectation value
$\log\langle W\rangle=-I_{\rm M2}$.
General geometric and semiclassical treatments are given in
\cite{Farquet:2013cwa,Farquet:2014bda,Gautason:2025bft}.
For a specified M-theory circle, its generator must
be proportional to the internal Reeb vector along
the wrapped orbit. This is the geometric form of
the critical-orbit condition in
\cite[Secs.~3.3, 3.4]{Farquet:2013cwa}.

Let $D_2\subset X_8$ be the invariant cone over such
a calibrated orbit. The auxiliary cycle is
$D_4=\BC_{\omega_a}\times D_2$, $a=1,2$, at the
origin of the other external line: the first factor
models the external disk, while $D_2$ fills the
internal circle. Using the same $E_4$ insertion,
$\omega_1\omega_2=1$ and
$C_{D_2}(-\epsilon)=-C_{D_2}(\epsilon)$ gives
\begin{equation}
 \begin{aligned}
 I_{\rm M2}(\mu)
 &=-\frac{\mu}{\omega_a}C_{D_2}(\epsilon),\\
 I_{\rm M2}(N)
 &=-\frac{2\pi\sqrt{2N}}{\omega_a}
       \frac{C_{D_2}(\epsilon)}{\sqrt{C_X(\epsilon)}}.
 \end{aligned}
 \label{eq:M2Wilson}
\end{equation}
The minus sign follows from the signed internal
curve pushforward; the background flux class \eqref{eq:M2AdS4background}
is unchanged.
Before imposing $\omega_1\omega_2=1$, the same calculation
gives $I_{\rm M2}(\mu)=-\mu\omega_{3-a}C_{D_2}(\epsilon)$,
with $a=1,2$.

For pure boundary squashing, the internal Reeb
direction is fixed and its weights rescale by
$(b+b^{-1})/2$ relative to the round-sphere
normalization. Since $C_{D_2}/\sqrt{C_X}$ is
homogeneous of degree one, the action acquires
the factor $(b+b^{-1})/(2\omega_a)$.
This agrees with the general Wilson-loop result
of \cite[Eqs.~(1.2), (1.3)]{Farquet:2014bda},
where the two polar circles have normalized
lengths $1/\omega_a$.

For ABJM, a calibrated coordinate orbit has
$D_2=\BC/\BZ_k$ and
$C_{D_2}=1/(k\epsilon_i)$.
At the pure-squashing locus
$\epsilon_i=(b+b^{-1})/4$, $\forall i$, the two actions are
\begin{equation}
 \begin{aligned}
 \left.I_{\rm M2}\right|_{a=1}
 &=-\frac{1+b^{-2}}{2}\,\pi\sqrt{\frac{2N}{k}},\\
 \left.I_{\rm M2}\right|_{a=2}
 &=-\frac{1+b^2}{2}\,\pi\sqrt{\frac{2N}{k}}.
 \end{aligned}
 \label{eq:ABJMWilson}
\end{equation}
They are exchanged by $b\leftrightarrow b^{-1}$.
At $b=1$, both reduce to
$I_{\rm M2}=-2\mu_*/k=-\pi\sqrt{2N/k}$.
This is the classical action of the
$\mathrm{AdS}_2\times S^1$ membrane dual to the
circular half-BPS Wilson loop in
\cite[Eq.~(2.9)]{Giombi:2023vzu}.~\footnote{With $R$ the internal seven-sphere radius,
the leading round-sphere saddle is
$\mu_*=R^3/(8\ell_p^3)$.
The direct calculation in
\cite[Eqs.~(2.8), (2.9)]{Giombi:2023vzu} gives
$-R^3/(4k\ell_p^3)=-2\mu_*/k$.
The one-loop factor computed there is separate
from the classical pairing considered here.}
These comparisons provide non-compact checks of
the volume prescription: it reproduces the
holographically renormalized actions with the
specified circular Wilson-loop boundary conditions. 

For $X_8=C(Q^{1,1,1})$, choose a unit-wrapped,
non-collapsed calibrated M-theory circle as in
\cite[Sec.~4.2]{Farquet:2013cwa}.~\footnote{Using spherical coordinates $(\theta_a,\varphi_a)$,
$a=1,2$, on the first two base spheres, the circle
is generated by
$\zeta_M=\partial_{\varphi_1}+\partial_{\varphi_2}$,
with contact Hamiltonian
$h_M=(\cos\theta_1+\cos\theta_2)/4$.
We choose the critical locus $h_M=1/2$ and a toric
representative in the remaining projective factor.}
Using homogeneous pairs $(x_1,x_2)$, $(y_3,y_4)$
and $(z_5,z_6)$, fixing a coordinate point in each
projective factor sets the other coordinate to zero. The resulting ray has invariant coordinate
$u_{ij\ell}=x_i y_j z_\ell$, with
$i=1,2$, $j=3,4$ and $\ell=5,6$, restricted to the
chosen non-collapsed locus.
Its weight is $\epsilon_i+\epsilon_j+\epsilon_\ell$,
so
$C_{D_2}=1/(\epsilon_i+\epsilon_j+\epsilon_\ell)$.
At the symmetric weights below \eqref{eq:Q111refined},
$C_X=6$ and $C_{D_2}=1$, giving
\begin{equation}
 I_{\rm M2}(N)=-2\pi\sqrt{\frac{N}{3}},
 \label{eq:Q111Wilson}
\end{equation}
in agreement with the $k=1$ result of
\cite[Eq.~(4.22)]{Farquet:2013cwa}.
Boundary squashing is incorporated by the factor
$(b+b^{-1})/(2\omega_a)$ derived above.

\subsubsection{Giant gravitons and defects in thermal
$\mathrm{AdS}_7\times S^4$}
\label{sec:M2AdS7}

For the six-dimensional index on $S^1\times S^5$,
the auxiliary background is
\cite{Hristov:2026tde,Hristov:2026uxs}
\begin{equation}
 \begin{aligned}
 Y_{12}&=\BC_{\epsilon_0}\times
 \BC^3_{\epsilon_1,\epsilon_2,\epsilon_3}\times S^4,\\
 E_4&=\frac N2e_T(TS^4).
 \end{aligned}
 \label{eq:M2AdS7background}
\end{equation}
Here $N$ is the M5 charge, with normalization fixed by
$\chi(S^4)=2$. The external factors extend the time circle and $S^5$,
with unit pushforward
$C_X=(\epsilon_0\epsilon_1\epsilon_2\epsilon_3)^{-1}$. We set $\epsilon_0=2\pi\ii$ and denote the
internal chemical potentials by $\Delta_1,\Delta_2$,
with signed auxiliary weights $-\Delta_1,-\Delta_2$.
Supersymmetry requires $\Delta_1+\Delta_2
=\epsilon_0+\epsilon_1+\epsilon_2+\epsilon_3$.
At leading order in $N$, the background action is
\begin{equation}
 I_{\rm M}^{(2\partial)}
 =\frac{N^3\Delta_1^2\Delta_2^2}
        {24\epsilon_1\epsilon_2\epsilon_3}.
 \label{eq:M2AdS7backgroundaction}
\end{equation}
At equal external weights
$\epsilon_0=\epsilon_1=\epsilon_2=\epsilon_3$, this is
proportional to the six-dimensional trial $a$-anomaly,
with $\Delta_{1,2}$ arbitrary subject to
the supersymmetric constraint above.~\footnote{This
comparison concerns the trial anomaly function.
For a stationary thermal background, the physical
circle integral supplies the period $\beta$,
giving $I_{\rm thermal}=\beta E_{\rm vac}$.
The vacuum energy requires the corresponding background
and boundary prescription.}

Choose the invariant $S^2\subset S^4$ represented by
the first coordinate line in the $\BC^2$ charts at
the two poles. The $\Delta_1$ rotation acts within
this sphere, while $\Delta_2$ rotates its normal plane.
Its calibrated orientation is opposite to the
coordinate-induced orientation in the signed convention.~\footnote{Conjugating both internal complex coordinates
preserves the orientation of $S^4$ but reverses that
of the coordinate two-sphere. The probe orientation
and charge are kept fixed.}
The M2 wrapping $S^1\times S^2$ is represented by
the oriented invariant subcycle~\footnote{
A radial truncation of the auxiliary line gives
a filling with boundary $S^1\times S^2$.
The physical time circle remains non-contractible
in thermal $\mathrm{AdS}_7$, with period distinct
from $\epsilon_0$.}
\begin{equation}
 D_4=\BC_{\epsilon_0}\times\{0\}\times S^2
 \subset Y_{12}.
 \label{eq:M2AdS7cycle}
\end{equation}

The splitting $TS^4|_{S^2}=TS^2\oplus NS^2$
fixes the normal orientation. For the calibrated
$S^2$, it gives $e_T(NS^2)=+\Delta_2$, so
multiplicativity of the Euler class yields
\begin{equation}
 \iota_D^*E_4=\frac N2\,\Delta_2\,e_T(TS^2).
 \label{eq:M2AdS7restriction}
\end{equation}
The auxiliary line contributes $1/\epsilon_0$,
while $\int_{S^2}e_T(TS^2)=2$. Thus
\eqref{eq:M2ambientpairing} gives
\begin{equation}
 I_{\rm M2}
 =\frac{2\pi\ii N\Delta_2}{\epsilon_0}
 =N\Delta_2.
 \label{eq:M2AdS7action}
\end{equation}

Equivalently, the calibrated cycle class is
$[D_4]_T=\epsilon_1\epsilon_2\epsilon_3\Delta_2$,
opposite to that of the coordinate zero locus.
In \eqref{eq:projectionformula}, this insertion
cancels the three spatial weights in $C_X$,
while $\int_{S^4}E_4=N$ reproduces the same result.
Exchanging the internal planes gives
$I_{\rm M2}=N\Delta_1$.

These maximal M2 giant gravitons carry charge $N$
under the rotation normal to the wrapped sphere.
Their classical factors $e^{-N\Delta_2}$ and
$e^{-N\Delta_1}$ agree with the refined
giant-graviton expansion
\cite[Eq.~(3.24)]{Arai:2020uwd}.
Their actions scale as $N$, compared with the
$N^3$ background action.

For comparison with \cite{Beccaria:2023sph},
we use the unrefined index specialization
\cite[Sec.~2.1]{Bullimore:2014upa}
\begin{equation}
 \epsilon_1=\beta,\qquad
 (\Delta_1,\Delta_2)=(2\beta,\beta).
 \label{eq:M2AdS7thermalweights}
\end{equation}
With Euclidean time $y\sim y+\beta$ and azimuthal
angle $\varphi$ in the normal two-plane to the giant,
the twist $\dd\varphi\mapsto\dd\varphi+\ii\dd y$
gives the normal complex coordinate the multiplier
$e^{-\beta}$ around the time circle, confirming
$\Delta_2=\beta$.
The remaining auxiliary weights satisfy
$\epsilon_2+\epsilon_3=2\beta-2\pi\ii$ and do not
enter these probe actions.

The $\Delta_2$ giant wraps $S^1_\beta\times S^2$
at the centre of $\mathrm{AdS}_7$.
Its volume and Wess--Zumino terms are $2N\beta$
and $-N\beta$, giving $I_{\rm M2}=N\beta$
in agreement with \eqref{eq:M2AdS7action}
\cite[Eqs.~(4.1)--(4.3)]{Beccaria:2023sph}.
The $\Delta_1$ family has classical action $2N\beta$,
but its index contribution vanishes in this Schur
specialization because of fermionic zero modes
\cite[Sec.~3.4]{Arai:2020uwd}.

A noncompact M2 instead wraps thermal
$\mathrm{AdS}_3\subset\mathrm{AdS}_7$ at an internal pole.
Its boundary $S^1_\beta\times S^1$ defines a Wilson
surface in the six-dimensional theory, reducing to a
supersymmetric Wilson loop in five-dimensional SYM
upon compactification on $S^1_\beta$
\cite{Beccaria:2023sph,Bullimore:2014upa}.
For the calibrated embedding, choose the pole $p_-$
with $e_T(TS^4)|_{p_-}=-\Delta_1\Delta_2$ and take
\begin{equation}
 D_4
 =\BC_{\epsilon_0}\times\BC_{\epsilon_1}
  \times\{0\}\times\{p_-\}\subset Y_{12},
 \label{eq:M2AdS7defectcycle}
\end{equation}
where $\{0\}$ fixes the other two spatial directions.
Since $E_4|_{p_-}=-N\Delta_1\Delta_2/2$, the same
linear pairing gives
\begin{equation}
 I_{\rm M2}
 =-\frac{\pi\ii N\Delta_1\Delta_2}{\epsilon_0\epsilon_1}
 =-\frac{N\Delta_1\Delta_2}{2\epsilon_1}.
 \label{eq:M2AdS7defectaction}
\end{equation}
At \eqref{eq:M2AdS7thermalweights}, this gives
$I_{\rm M2}=-N\beta$ and classical weight $e^{N\beta}$,
matching the renormalized action of
\cite[Eqs.~(5.1)--(5.3)]{Beccaria:2023sph}.
The refined expression has the same leading parameter
dependence as the fundamental-defect Casimir energy
\cite[Eq.~(25), $k=1$]{Huang:2025eze}, with
$\Delta_A=\beta\sigma_A$ and
$\epsilon_1=\beta\omega_1$.

\section{D3-branes}
\label{sec:D3}

We consider type IIB backgrounds with constant axion
and dilaton and only five-form flux. The auxiliary
geometry is
\begin{equation}
 Y_{12}=X_6\times Z_6,\qquad Z_6=C(SE_5),
 \label{eq:D3backgrounddata}
\end{equation}
where the factors model equivariant fillings of the
external spacetime $\mathcal M_5$ and the reference
internal Sasaki--Einstein space $SE_5$.
Their corner models the physical ten-dimensional geometry.
We assume compatible extensions of the torus action
and potential data, and factorization of the completed
relative pairings as specified below.
The auxiliary flux model is restricted to mesonic
compactification data.~\footnote{For the explicit
contact derivation below, we use a fixed Sasaki--Einstein
fibre, fibred over spacetime only along its Reeb direction,
with the corresponding five-form uplift
\cite{Colombo:2025ihp,BenettiGenolini:2026hmz}.
This corresponds to minimal gauged supergravity in
five dimensions. Our proposed auxiliary framework
allows more general mesonic fibrations and associated
metric and flux deformations, subject to the matching
assumptions below.}

For external and internal equivariant parameters
$\epsilon_i$ and $\omega_\alpha$, respectively, we use
signed auxiliary weights $(\epsilon;-\omega)$ and impose
\cite{Hristov:2026tde,Hristov:2026uxs}
\begin{equation}
 c_1^T(TY_{12})
 =\sum_i\epsilon_i-\sum_\alpha\omega_\alpha=0.
 \label{eq:D3CYcondition}
\end{equation}
We write $C_X=C_X(\epsilon)$ and $C_Z=C_Z(\omega)$
for the unit pushforwards of
appendix~\ref{app:equivariant}, so $C_Z(-\omega)=-C_Z$.
The internal sign reversal also applies to
characteristic classes and cycle volumes.
Physical orientations and brane charges are specified
separately.

\subsection{Type IIB action}
\label{sec:IIBaction}

We choose the ensemble with fixed magnetic five-form
flux. In the five-form sector with constant axion and
dilaton, the standard bulk pseudo-action vanishes on
shell. For separated electric and magnetic flux,
the fixed-flux variational principle requires the
additional topological term
$-\ii(4\kappa_{10}^2)^{-1}
\int F_5^{\rm el}\wedge F_5^{\rm mag}$
\cite{Kurlyand:2022vzv,vanMuiden:2026lno}.
Using $\star_{10}F_5=\ii F_5$, the resulting
two-derivative on-shell action is
\begin{equation}
 I_{\rm IIB}^{\rm mag}
 =\frac{1}{4\kappa_{10}^2}
 \int F_5^{\rm mag}\wedge\star_{10}F_5^{\rm mag}+I_\partial,
 \label{eq:IIBnormsquares}
\end{equation}
where $2\kappa_{10}^2=(2\pi)^7\ell_s^8$ and
$I_\partial$ contains the Gibbons--Hawking--York
term and counterterms.

We illustrate the construction for $SE_5$ fibred over
$\mathcal M_5$, using the contact representative of
\cite{Colombo:2025ihp,BenettiGenolini:2026hmz}.
Let $\sigma$ be the external supersymmetric one-form,
including the gauge potential, and $\eta$ the contact
form of $SE_5$, both in unit-radius conventions.
The supersymmetric vector
$K^M=\frac12\bar\varepsilon\Gamma^M\varepsilon$,
defined by the IIB Killing-spinor doublet $\varepsilon$,
acts through its projection to each factor.
Assume invariant representatives satisfying
\begin{equation}
 \iota_K\sigma=\iota_K\eta=c\ne0,\qquad
 \iota_K\dd\sigma=\iota_K\dd\eta=0.
 \label{eq:D3commonK}
\end{equation}
Here $c$ is an ordinary scalar, linear in the
equivariant parameters $\epsilon_i$ and $\omega_\alpha$.
These parameters have degree two in the Cartan grading,
so $c$ has equivariant degree two and ordinary form
degree zero. We normalize the internal flux by
\begin{equation}
 \frac{\left.F_5\right|_{SE_5}}{(2\pi\ell_s)^4}
 =q\,\eta\wedge(\dd\eta)^2,\qquad
 q\int_{SE_5}\eta\wedge(\dd\eta)^2=N.
 \label{eq:D3fluxnormalization}
\end{equation}
Using the reduced supersymmetric Lagrangian of
\cite[Eqs.~(3.34)--(3.37)]{Colombo:2025ihp}
and $I=-\ii S$ for the Lorentzian action $S$,
\begin{equation}
 I_{\rm IIB}^{\rm mag}
 =\pi\ii q^2
 \left(\int_{\mathcal M_5}\sigma\wedge(\dd\sigma)^2\right)
\left(\int_{SE_5}\eta\wedge(\dd\eta)^2\right)+I_{\rm bdy},
 \label{eq:IIBcontactaction}
\end{equation}
where $I_{\rm bdy}$ includes $I_\partial$, the integrated
exact term and gauge-patching contributions.

For $\chi=\sigma,\eta$, define
$\Theta_\chi:=\chi/(c-\dd\chi)$ by the finite expansion
in \eqref{eq:relativeunit}. Since
$\dd_K\Theta_\chi=-1$, the pair $(1,\Theta_\chi)$
represents the relative unit on a compatible filling.
The same primitive supplies an algebraic relative
lift with several vector multiplets.~\footnote{The five-dimensional relative insertion is proportional
to $P_6=\frac16C_{IJK}\Phi^I\wedge\Phi^J\wedge\Phi^K$,
where $C_{IJK}$ are the symmetric Chern--Simons couplings
and the equivariantly closed polyforms
$\Phi^I=\mathcal F^I+\Phi_0^I$ combine gauge curvatures
$\mathcal F^I$ with scalar equivariant components
$\Phi_0^I$
\cite{BenettiGenolini:2026cdw,Colombo:2025yqy,Cassani:2026teb}.
For a boundary inclusion $j$,
$\dd_K(j^*P_6\,\Theta_\sigma)=-j^*P_6$.
Thus one primitive supplies the boundary representative
while retaining all vector data in $P_6$.
}
We encode the normalized internal flux by the relative pair
\begin{equation}
 (E_6^{\rm mag},E_5^{\rm mag})_{\rm phys}
 :=\bigl(qc^3,qc^3\Theta_\eta\bigr),
 \label{eq:IIBfluxdictionary}
\end{equation}
such that
$E_5^{\rm mag}
=q\,\eta\wedge[(\dd\eta)^2+c\,\dd\eta+c^2]$ is a polyform of degree five with top form matching the physical flux in \eqref{eq:D3fluxnormalization}. Its lower-degree components ensure
\begin{equation}
 \dd_K E_6^{\rm mag}=0,\qquad
 j^*E_6^{\rm mag}=-\dd_K E_5^{\rm mag},
 \label{eq:IIBdescent}
\end{equation}
where $j$ includes $SE_5$ into its six-dimensional
filling.
The subscripts denote total equivariant degree:
$E_6^{\rm mag}$ is an ordinary scalar of degree six,
while $E_5^{\rm mag}$ combines ordinary forms of degrees
one, three and five into total degree five.

The relative product of the external unit
$(1,\Theta_\sigma)$ and the internal pair
$(\frac12(E_6^{\rm mag})^2,
\frac12(E_6^{\rm mag})^2\Theta_\eta)$
has bulk member $\frac12(E_6^{\rm mag})^2$ and corner
density
\begin{equation}
 \left[\frac12(E_6^{\rm mag})^2\Theta_\sigma\wedge\Theta_\eta\right]_{10}
 =\frac12q^2\,\sigma\wedge(\dd\sigma)^2\wedge\eta\wedge(\dd\eta)^2.
 \label{eq:IIBsquareunit}
\end{equation}
With the orientations of appendix~\ref{app:equivariant},
its integral multiplied by $2\pi\ii$ reproduces the
contact term in \eqref{eq:IIBcontactaction}.

We assume that the auxiliary model $Y_{12}=X_6\times Z_6$
preserves the complete relative pairing, including
$I_{\rm bdy}$.
Its flux class, again denoted $E_6^{\rm mag}$,
must reproduce the reduced mesonic insertion through
its internal quadratic pushforward
\cite{Benvenuti:2006xg,Bah:2020jas}.
Then
\begin{equation}
\begin{split}
 I_{\rm IIB}^{\rm mag}
 =2\pi\ii\left(\int_{X_6}^{\mathbb V}1\right)
 &\left(\int_{Z_6}^{\mathbb V}\frac12(E_6^{\rm mag})^2\right)\\
 =&2\pi\ii\int_{Y_{12}}^{\mathbb V}\frac12(E_6^{\rm mag})^2,
 \label{eq:IIBbackgroundmatching}
\end{split}
\end{equation}
in normalized Chern-root conventions with internal
weights $-\omega$, exhibiting the quadratic structure
of anomaly inflow
\cite{Bah:2020jas,Hristov:2026uxs}.

In what follows, we choose the auxiliary flux class proportional to
the equivariant unit:
\begin{equation}
 \int_{Z_6}^{\mathbb V}E_6^{\rm mag}=N,\qquad
 E_6^{\rm mag}=\frac{N\,1_{Z_6}}{C_Z(-\omega)}
             =-\frac{N\,1_{Z_6}}{C_Z}.
 \label{eq:IIBmagneticnormalization}
\end{equation}
This defines a degree-six class in localized
equivariant cohomology.
Independent baryonic data require additional class
components, which can have vanishing total pushforward
while contributing to quadratic or probe pairings.
For $Z_6=\BC^3$, the quadratic pushforward reproduces
the STU cubic \cite{Cvetic:1999xp}.
For general toric cones, the inverse-volume relation
reproduces the cubic anomaly after extremizing over
baryonic potentials at fixed mesonic weights
\cite{Benvenuti:2006xg,Butti:2005vn}.
Under the matching assumption above, the action is
\begin{equation}
 I_{\rm IIB}^{\rm mag}
 =-\pi\ii N^2\frac{C_X}{C_Z}.
 \label{eq:IIBmagneticaction}
\end{equation}

\subsection{Probe D3 action}
\label{sec:D3action}

Within the contact specialization above, consider a
supersymmetric worldvolume with compatible oriented
equivariant product fillings,
\begin{equation}
 \begin{aligned}
 W_6&=W_{2m}^{X}\times W_{2n}^{Z},\\
 \Sigma_4&=-\,\Sigma_{2m-1}^{\rm ext}
              \times\Sigma_{2n-1}^{\rm int},
 \qquad (m,n)=(1,2),(2,1).
 \end{aligned}
 \label{eq:D3probefillings}
\end{equation}
The worldvolume factors lie in $\mathcal M_5$ and $SE_5$,
with $\partial W_{2m}^{X}=\Sigma_{2m-1}^{\rm ext}$
and $\partial W_{2n}^{Z}=\Sigma_{2n-1}^{\rm int}$
carrying the induced boundary orientations.
The displayed corner orientation must agree with the
calibrated brane orientation.
We model the physical filling $W_6$ by
$D_6=D_X\times D_Z\subset Y_{12}$ with its product
orientation.

With vanishing gauge-invariant worldvolume two-form,
the Euclidean action is
\cite{Cederwall:1996pv,Bergshoeff:1996tu}
\begin{equation}
 I_{\rm D3}^{\rm mag}
 =\mu_3\int_{\Sigma_4}\vol_{\Sigma_4}
 -\ii\mu_3\int_{\Sigma_4}P[C_4],
 \label{eq:D3action}
\end{equation}
where $\mu_3=[(2\pi)^3\ell_s^4]^{-1}$,
$C_4$ is the Ramond--Ramond potential with
$F_5=\dd C_4$ locally, and $P$ denotes pullback.
The potential, embedding, spinors and Einstein-frame
volume branch are continued together from Lorentzian
signature.
The $\kappa$-symmetry condition makes $K$ tangent
to $\Sigma_4$
\cite{Martucci:2011dn}.~\footnote{With the charge and
orientation conventions of \eqref{eq:D3action}, define
\begin{equation}
 \Phi_3=-\frac{\ii}{2}
 \bar\varepsilon\Gamma_{(3)}\tau_2\varepsilon,
 \label{eq:D3bilinears}
\end{equation}
where $\tau_2$ is the second Pauli matrix acting on
the IIB Killing-spinor doublet.
Analytically continuing the Lorentzian D3 projector
\cite[Eqs.~(4.3)--(4.4)]{Martucci:2011dn}
in our conventions gives
\begin{equation}
\Gamma_{\Sigma_4}\tau_2\varepsilon=\varepsilon
 \quad\Rightarrow\quad
 \iota_K\vol_{\Sigma_4}=\ii P[\Phi_3],
 \label{eq:D3calibration}
\end{equation}
where $\Gamma_{\Sigma_4}$ is the normalized Clifford
volume element. The calibration identity follows
as in footnote~\ref{ft:M2}.}

The calibration and Wess--Zumino term evaluated in
\cite{BenettiGenolini:2026hmz} give, in our conventions,
\begin{equation}
 I_{\rm D3}^{\rm mag}
 =-2\pi\ii q\int_{\Sigma_4}
       \sigma\wedge\eta\wedge(\dd\sigma+\dd\eta),
 \label{eq:D3contactaction}
\end{equation}
using $\mu_3(2\pi\ell_s)^4=2\pi$.
Here $\sigma$ and $\eta$ are the same background
one-forms as in the previous subsection, with their
pullbacks to the external and internal worldvolume
factors understood. Restrict the external relative unit $(1,\Theta_\sigma)$
and the internal flux pair
$(E_6^{\rm mag},E_6^{\rm mag}\Theta_\eta)$
to the respective fillings.
Their product has bulk member $E_6^{\rm mag}$ and
corner density
\begin{equation}
 \left[E_6^{\rm mag}\Theta_\sigma\wedge\Theta_\eta\right]_4
 =q\,\sigma\wedge\eta\wedge(\dd\sigma+\dd\eta),
 \label{eq:D3Stokes}
\end{equation}
where $E_6^{\rm mag}=qc^3$ is the physical contact
representative of \eqref{eq:IIBfluxdictionary}.
Only the $\dd\eta$ term survives for $(m,n)=(1,2)$,
and only the $\dd\sigma$ term for $(m,n)=(2,1)$.
Equation~\eqref{eq:D3commonK} with $c\ne0$ ensures that the
components of $K$ along both worldvolume factors are nowhere zero.

For compact fillings, the relative product formula
of appendix~\ref{app:equivariant} gives
\begin{equation}
 \begin{aligned}
 \int_{W_6}^{\rm eq}E_6^{\rm mag}
 &=\int_{\Sigma^{\rm ext}\times\Sigma^{\rm int}}
       [E_6^{\rm mag}\Theta_\sigma\wedge\Theta_\eta]_4\\
 &=-\int_{\Sigma_4}
       [E_6^{\rm mag}\Theta_\sigma\wedge\Theta_\eta]_4.
 \end{aligned}
 \label{eq:D3relativeorientation}
\end{equation}
The second equality uses \eqref{eq:D3probefillings}.
Combining this with \eqref{eq:D3contactaction} and
\eqref{eq:D3Stokes} yields
$I_{\rm D3}^{\rm mag}
=2\pi\ii\int_{W_6}^{\rm eq}E_6^{\rm mag}$.
The probe thus inserts linearly the same flux class
whose square enters the background action.

Assuming that the auxiliary model preserves this
calibrated pairing, including any required boundary
terms, we obtain
\begin{equation}
 I_{\rm D3}^{\rm mag}
 =2\pi\ii\int_{D_6}^{\mathbb V}\iota_D^*E_6^{\rm mag}
 =2\pi\ii\int_{Y_{12}}^{\mathbb V}E_6^{\rm mag}[D_6]_T,
 \label{eq:D3magneticprobe}
\end{equation}
where $\iota_D:D_6\hookrightarrow Y_{12}$ and
$[D_6]_T=(\iota_D)_*1$.
On the mesonic locus, \eqref{eq:IIBmagneticnormalization}
then gives
\begin{equation}
 I_{\rm D3}^{\rm mag}
 =-2\pi\ii N\,
 \frac{C_{D_X}(\epsilon)C_{D_Z}(-\omega)}{C_Z(\omega)},
 \label{eq:D3fixedcharge}
\end{equation}
with $C_{D_Z}(-\omega)=(-1)^nC_{D_Z}(\omega)$
for the conical models considered here.

\subsection{Electric ensemble}
\label{sec:D3ensembles}

We define $E_6^{\rm el}$ as an auxiliary representative
of the ensemble dual to the magnetic construction
\cite{Hristov:2026uxs,vanMuiden:2026lno}.
We normalize the chemical potential $\mu$ by
\begin{equation}
 I_{\rm IIB}^{\rm mag}(N)
 =\operatorname*{ext}_{\mu}
   \left[I_{\rm IIB}^{\rm el}(\mu)-2\pi\ii\mu N\right]\,.
 \label{eq:IIBpolarizations}
\end{equation}
For $X_6=\BC^3$, take
\begin{equation}
 E_6^{\rm el}=\ii\mu\,e_T(TX_6)\,,\qquad
 \int_{X_6}^{\mathbb V}E_6^{\rm el}=\ii\mu\,.
 \label{eq:IIBrepresentatives}
\end{equation}
The internal cone remains general. The quadratic pairing gives
\begin{equation}
 \begin{aligned}
 I_{\rm IIB}^{\rm el}
 =2\pi\ii\int_{Y_{12}}^{\mathbb V}
                      \frac12(E_6^{\rm el})^2=\pi\ii\mu^2\frac{C_Z}{C_X}\,.
 \end{aligned}
 \label{eq:IIBensembleactions}
\end{equation}
The Legendre transform reproduces
\eqref{eq:IIBmagneticaction}, with
\begin{equation}
 \mu_*=\frac{NC_X}{C_Z}\,,\qquad
 E_6^{\rm el}(\mu_*)=-\ii E_6^{\rm mag}\,,
 \label{eq:IIBensembles}
\end{equation}
where the second equality concerns the scalar auxiliary
representatives used here.

The probe insertion is included in the ensemble
transform described above. At leading order, stationarity of the
background saddle gives
$I_{\rm D3}^{\rm mag}(N)=I_{\rm D3}^{\rm el}(\mu_*)$.
The corresponding linear representative is
\begin{equation}
 \begin{aligned}
 I_{\rm D3}^{\rm el}
 =-2\pi\int_{D_6}^{\mathbb V}\iota_D^*E_6^{\rm el}=-2\pi\int_{Y_{12}}^{\mathbb V}E_6^{\rm el}[D_6]_T.
 \end{aligned}
 \label{eq:D3electricprobe}
\end{equation}
By \eqref{eq:IIBensembles}, this equals
\eqref{eq:D3magneticprobe} at the saddle, reproducing
\eqref{eq:D3fixedcharge}.

\subsection{Examples}
\label{sec:D3examples}

We use the magnetic representation at fixed background
charge $N$ and suppress the superscript.
Keeping the internal mesonic weights $\omega$ general,
we take
\begin{equation}
 X_6=\BC_{\epsilon_0}\times
      \BC^2_{\epsilon_1,\epsilon_2},
 \qquad
 C_X=\frac1{\epsilon_0\epsilon_1\epsilon_2}.
 \label{eq:D3external}
\end{equation}
The factor $\BC_{\epsilon_0}$ models the thermal circle,
with $\epsilon_0$ specified for each background, while
$\epsilon_1,\epsilon_2$ are the external angular
chemical potentials.
On the Reeb specialization specified below, the compact
black hole probes and surface defects reproduce
\cite{BenettiGenolini:2026hmz}, whose results already
cover arbitrary toric $SE_5$. For these external data, \eqref{eq:IIBmagneticaction}
gives
\begin{equation}
 I_{\rm IIB}
 =-\frac{\pi\ii N^2}
 {\epsilon_0\epsilon_1\epsilon_2 C_Z(\omega)}.
 \label{eq:D3backgroundaction}
\end{equation}
At $\epsilon_0=\epsilon_1=\epsilon_2$, this is
proportional to the four-dimensional trial $a$-anomaly
on the mesonic locus \cite{Hristov:2025ygn}, with internal weights subject to
\eqref{eq:D3CYcondition}.

\subsubsection{Compact black hole instantons}
\label{sec:D3blackholes}

For asymptotically $\mathrm{AdS}_5\times SE_5$
black holes, we use $\epsilon_0=-2\pi\ii$.
The thermal orbit, including the angular shifts
of the rotating saddle, contracts at the horizon.
This contractibility is supplied by the physical
geometry in addition to the auxiliary equivariant data.

We consider two compact probe families.
First, let $\Sigma_I\subset SE_5$ be a calibrated
three-cycle with surface cone $D_I^{(4)}\subset Z_6$.
A probe wraps $L_a\times\Sigma_I$, where $L_a$ is
one of the two distinguished toric circles on the
horizon $S^3$.~\footnote{Although $L_a$ is contractible in the horizon
$S^3$, the full D3 embedding satisfies the
$\kappa$-symmetry condition
\cite[App.~C.3]{Aharony:2021zkr}.
Contracting the circle need not preserve this condition.} Its auxiliary model is
$D_6=\BC_{\epsilon_a}\times D_I^{(4)}$, with
$C_{D_X}(\epsilon)=1/\epsilon_a$, $a=1,2$. We find from \eqref{eq:D3fixedcharge}
\begin{equation}
 I_{\rm D3}
 =-2\pi\ii N
 \frac{C_{D_I^{(4)}}(\omega)}
      {\epsilon_a C_Z(\omega)}\,,
 \qquad a=1,2\,.
 \label{eq:D3internalaction}
\end{equation}

Second, a probe can wrap the horizon $S^3$ and
an internal closed toric orbit with curve cone
$D_I^{(2)}$.
Its auxiliary cycle is
$D_6=\BC^2_{\epsilon_1,\epsilon_2}\times D_I^{(2)}$.
Using $C_{D_I^{(2)}}(-\omega)
=-C_{D_I^{(2)}}(\omega)$ in
\eqref{eq:D3fixedcharge} gives
\begin{equation}
 I_{\rm D3}
 =2\pi\ii N
 \frac{C_{D_I^{(2)}}(\omega)}
      {\epsilon_1\epsilon_2 C_Z(\omega)}\,.
 \label{eq:D3horizonaction}
\end{equation}

A direct comparison is available on the Reeb
specialization
\begin{equation}
 \omega=\varphi b,\qquad
 3\varphi=\epsilon_1+\epsilon_2\mp2\pi\ii,
 \label{eq:D3universallocus}
\end{equation}
where $b$ is the Sasaki--Einstein Reeb vector,
normalized to give the holomorphic three-form weight
three, and $\varphi$ is the R-symmetry chemical potential.
The upper/lower signs label the first/second BPS
branches, with supersymmetric anti-D3/D3 charge
for the horizon-wrapping probe, respectively
\cite[App.~C.3.3]{Aharony:2021zkr}.
We use the first branch. With $I=-\ii S$, our two
probe families reproduce Eq.~(31) and the anti-D3
version of Eq.~(34) in \cite{BenettiGenolini:2026hmz}. On this locus, the background action becomes
$2a_{\rm CFT}\varphi^3/(\epsilon_1\epsilon_2)$,
with central charge $a_{\rm CFT}=N^2/[4C_Z(b)]$, reproducing
\cite[Eq.~(30)]{BenettiGenolini:2026hmz}.

We now evaluate the same prescription at general
mesonic weights, beyond the Reeb specialization.

For $SE_5=S^5$, the cone $Z_6=\BC^3$ has coordinates
$z_i$ of weights $\omega_i$, with
$C_Z=1/(\omega_1\omega_2\omega_3)$ and
$\sum_i\omega_i=\epsilon_1+\epsilon_2-2\pi\ii$.
Choosing the surface $z_i=0$ for the first family
and the $z_i$ axis for the second gives
\begin{align}
 I_{\rm D3}
 &=-2\pi\ii N\frac{\omega_i}{\epsilon_a}\,, \qquad \qquad a=1, 2,
 \label{eq:D3S5internal}\\
 I_{\rm D3}
 &=2\pi\ii N
   \frac{\omega_j\omega_k}{\epsilon_1\epsilon_2}\,,
 \qquad \{i,j,k\}=\{1,2,3\}.
 \label{eq:D3S5horizon}
\end{align}
At $\omega_i=\varphi$, these recover
\cite[Eqs.~(32), (35)]{BenettiGenolini:2026hmz}
and \cite[App.~C]{Aharony:2021zkr},
with the action and charge conventions above.
The first family also matches the Bethe Ansatz
exponents at $\epsilon_1=\epsilon_2$
\cite[Eq.~(3.53)]{Aharony:2021zkr}.~\footnote{
We use
$(\epsilon_1,\epsilon_2,\omega_i)
=2\pi\ii(\sigma_g,\tau_g,\Delta_{g,i})$,
the basic saddle $m=1$, $r=0$, and first-chamber
internal potentials.}
The second agrees with the gravitational conjecture
in \cite[Eq.~(D.35)]{Aharony:2021zkr}, supported by
analytic special cases and numerical checks.
Its identification with a specific field-theory
index sector remains open.

For $SE_5=T^{1,1}$, dual to Klebanov--Witten theory
\cite{Klebanov:1998hh}, the conifold $Z_6=C(T^{1,1})$
admits the small resolution
$\mathcal O(-1)\oplus\mathcal O(-1)\to\mathbb P^1$.
The homogeneous coordinates $(A_1,A_2,B_1,B_2)$ have
charges $(1,1,-1,-1)$ and weights $\omega_I$.
Writing $z_{ij}=A_iB_j$, set
$w_{ij}:=\omega_i+\omega_{j+2}$, with $i,j=1,2$, and
$w_{\rm tot}:=\sum_I\omega_I$.
At vanishing resolution parameter, the equivariant
volume is \cite{Martelli:2023oqk,Cassia:2025aus}
\begin{equation}
 C_Z=\frac{w_{\rm tot}}{\prod_{i,j=1}^2w_{ij}},
 \qquad w_{\rm tot}=\epsilon_1+\epsilon_2-2\pi\ii.
 \label{eq:conifoldvolume}
\end{equation}
The mesonic representative
$\omega_1\omega_2=\omega_3\omega_4$
\cite[Sec.~4.2]{Cassia:2025aus} fixes the redundant
homogeneous weights while retaining the general
effective torus action.

Supersymmetric probe embeddings were studied in
\cite{Arean:2004mm}.
On the cone, $A_i=0$ and $B_j=0$ define $\BC^2$
surfaces. Their volume ratios on the mesonic
representative are
$C_{D_Z(A_i=0)}/C_Z=\omega_i$ and
$C_{D_Z(B_j=0)}/C_Z=\omega_{j+2}$.
Taking $D_X=\BC_{\epsilon_a}$ gives
\begin{equation}
 \begin{aligned}
 I_{\rm D3}(A_i=0)
 &=-2\pi\ii N\frac{\omega_i}{\epsilon_a}\,,\\
 I_{\rm D3}(B_j=0)
 &=-2\pi\ii N\frac{\omega_{j+2}}{\epsilon_a}\,.
 \end{aligned}
 \label{eq:D3conifoldinternal}
\end{equation}
For the horizon-wrapping family, the curve with only
$z_{ij}$ nonzero has volume $1/w_{ij}$, yielding
\begin{equation}
 I_{\rm D3}=2\pi\ii N\frac{w_{\rm tot}\,\omega_1\omega_2}{\epsilon_1\epsilon_2(\omega_i+\omega_{j+2})}\,.
 \label{eq:D3conifoldhorizon}
\end{equation}
At $\omega_I=3\varphi/4$, the two families reduce to
$-3\pi\ii N\varphi/(2\epsilon_a)$ and
$9\pi\ii N\varphi^2/(4\epsilon_1\epsilon_2)$,
respectively, reproducing
\cite[Eqs.~(33), (36)]{BenettiGenolini:2026hmz}
with the same charge conventions.
Their general mesonic dependence remains to be
checked independently in field theory.

\subsubsection{Non-compact black hole defects}
\label{sec:D3defects}

A D3-brane wrapping the external cigar, a spatial
circle and an internal closed toric orbit with cone
$D_I^{(2)}$ defines a surface defect supported on
the thermal circle times a circle in the boundary
$S^3$. Its auxiliary cycle is
\begin{equation}
 D_6=\BC_{\epsilon_0}\times
       \BC_{\epsilon_a}\times D_I^{(2)}.
 \label{eq:D3defectcycle}
\end{equation}
With $\epsilon_0=-2\pi\ii$,
\eqref{eq:D3fixedcharge} gives
\begin{equation}
 I_{\rm D3}
 =-N\frac{C_{D_I^{(2)}}(\omega)}
           {\epsilon_a C_Z(\omega)}.
 \label{eq:D3defectaction}
\end{equation}
On the Reeb locus $\omega=\varphi b$, this reproduces
\cite[Eq.~(39)]{BenettiGenolini:2026hmz}.
For $S^5$ and $T^{1,1}$, the respective specializations
$\omega_i=\varphi$ and $w_{ij}=3\varphi/2$ give
\begin{equation}
 I_{\rm D3}^{S^5}=-\frac{N\varphi^2}{\epsilon_a},
 \qquad
 I_{\rm D3}^{T^{1,1}}
 =-\frac{9N\varphi^2}{8\epsilon_a},
\end{equation}
recovering its equations~(42), (43).
The $S^5$ result also matches the independent gravity
and large-$N$ defect-index calculations
\cite[Eqs.~(2.37), (3.43)]{Chen:2023lzq}.

These comparisons subtract the thermal AdS probe
action from the black hole probe action at matched
boundary data, testing our volume regulator against
this finite prescription.
For the defect operator $\mathcal D$,
$\log\langle\mathcal D\rangle=-I_{\rm D3}$
at leading order in $N$.
The general mesonic dependence in
\eqref{eq:D3defectaction} refines these results.

\subsubsection{Giant gravitons in thermal
$\mathrm{AdS}_5\times SE_5$}
\label{sec:D3giants}

The supersymmetric thermal quotient of global
$\mathrm{AdS}_5$ has topology $S^1\times\mathbb R^4$.
We consider maximal giant gravitons at its centre,
wrapping the thermal circle and a calibrated
three-cycle $\Sigma_I\subset SE_5$ in co-rotating
coordinates
\cite{McGreevy:2000cw,Mikhailov:2000ya,Beasley:2002xv,GonzalezLezcano:2026wje}.
Homologically nontrivial internal cycles also
carry baryonic charge \cite{Gubser:1998fp}.

The auxiliary cycle is
\begin{equation}
 \epsilon_0=2\pi\ii,\qquad
 D_6=\BC_{\epsilon_0}\times D_I^{(4)},
 \label{eq:D3thermaldata}
\end{equation}
at the origin of the external $\BC^2$.
The physical thermal circle is non-contractible,
and \eqref{eq:D3CYcondition} fixes the internal
holomorphic three-form weight to
$\epsilon_1+\epsilon_2+2\pi\ii$.
The probe prescription \eqref{eq:D3fixedcharge} gives
\begin{equation}
 I_{\rm D3}
 =-N\,\frac{C_{D_I^{(4)}}(\omega)}{C_Z(\omega)}.
 \label{eq:D3giantaction}
\end{equation}
This differs from \eqref{eq:D3internalaction}
by the choice of external circle.
On the Reeb locus $\omega=\varphi b$,
$I_{\rm D3}=-\varphi E_{\rm BPS}$, where
$E_{\rm BPS}=N C_{D_I^{(4)}}(b)/C_Z(b)$
is the wrapped-brane energy in unit-radius conventions.

For $S^5$, the surfaces $z_i=0$ in $Z_6=\BC^3$ give
\begin{equation}
 I_{\rm D3}^{S^5}=-N\omega_i,\qquad
 e^{-I_{\rm D3}^{S^5}}=e^{N\omega_i},
 \qquad i=1,2,3.
\end{equation}
These reproduce the refined maximal-giant factors,
with independent internal weights subject to the
index constraint
\cite{Gaiotto:2021xce,Deddo:2025lfm},
and their half-BPS specialization
\cite{Eleftheriou:2023jxr,Eleftheriou:2025lac}.

For $T^{1,1}$, label the four toric divisors
$(A_1=0,A_2=0,B_1=0,B_2=0)$ by $I=1,\ldots,4$,
in the same order as their homogeneous weights
$\omega_I$. These correspond to the four vertices
of the conifold toric diagram.
Using the divisor-volume ratios above, the mesonic
representative $\omega_1\omega_2=\omega_3\omega_4$
gives
\begin{equation}
 I_{\rm D3}^{T^{1,1}}=-N\omega_I,\qquad
 e^{-I_{\rm D3}^{T^{1,1}}}=e^{N\omega_I},
 \qquad I=1,\ldots,4.
 \label{eq:D3giantconifoldpotentials}
\end{equation}
Here $\sum_I\omega_I=\epsilon_1+\epsilon_2+2\pi\ii$.
These exponential weights match
\cite[Eq.~(3.6)]{Arai:2019aou}
and \cite{Deddo:2026exw} on this mesonic locus.
At $\omega_I=3\varphi/4$, all four actions become
$-3N\varphi/4$, with BPS energy $3N/4$,
as in the wrapped-D3/dibaryon comparison
\cite{Gubser:1998fp} and the direct calculation
in \cite[Sec.~3.1]{Deddo:2026exw}.

\section{Outlook}
\label{sec:outlook}

Relations between M2- and M5-brane indices motivate
comparing distinct non-perturbative sectors within our auxiliary
framework. Exchanging worldvolume and transverse
geometries, together with ensembles, has been shown to relate perturbative indices, including controlled
higher-derivative terms, within the approximations
of \cite{Hristov:2026tde}, and admits a
twelve-dimensional interpretation \cite{Hristov:2026uxs}.
Giant-graviton expansions offer a complementary
perspective: M5 giants on internal five-spheres
contribute to the three-dimensional index, and M2
giants on internal two-spheres to the six-dimensional
index \cite{Arai:2020uwd,Imamura:2022aua}.
Their coefficients involve worldvolume indices
with transformed fugacities, yielding reciprocal
relations in protected limits \cite{Hayashi:2024aaf}
and motivating a proposed full-index correspondence
\cite{Chen:2026fpe}. Connections between D3- and M5-giant expansions and
master volumes at large wrapping numbers were developed
in \cite{Chen:2024erz}.
For sphere Wilson loops, the analytic continuations
of probe configurations in
\cite[Sec.~6]{Beccaria:2023sph} would need to be
supplemented by a matching of observables and
ensembles \cite{Gautason:2025plx}.

This motivates extending the construction to M5-brane
probes. The descent
of the dual M-theory flux suggests
\cite{Gauntlett:2002fz,BenettiGenolini:2026cdw}
\begin{equation}
 I_{\rm M5}\stackrel{?}{=}
 2\pi\ii\int_{D_8}^{\mathbb V}
       \frac12\iota_D^*(E_4^2),
 \qquad D_8\subset Y_{12},
 \label{eq:M5proposal}
\end{equation}
where $D_8$ would model a relative filling with the
six-dimensional worldvolume as a corner.
The proposed thermal extension of
\eqref{eq:M2AdS4background} gives $I_{\rm M5}=N\epsilon_i$
for maximal M5 giants wrapping $S^1\times S^5$ in
$\mathrm{AdS}_4\times S^7$, where $\epsilon_i$ is the
weight of the normal complex plane, consistent with
the known classical weights
\cite{Beccaria:2023cuo,Chen:2026fpe}.~\footnote{Take
$X_8=\BC^4$, $\omega_1=2\pi\ii$ and
$D_8=\BC_{\omega_1}\times\{0\}\times\{z_i=0\}$,
with internal weights $-\epsilon_j$.
Writing $C_X=1/\prod_j\epsilon_j$, the fixed-charge
relation $N=\mu^2(\omega_1\omega_2)^2C_X/(8\pi^2)$
gives $\tfrac12E_4^2=-N/C_X$.
Together with $C_{D_8}=-\epsilon_i C_X/\omega_1$,
this yields the stated action from
\eqref{eq:M5proposal}.}
This is an algebraic consistency check.
A derivation must establish the relative data and
normalization from the calibrated M5 action and
determine how the field strength $H_3=\dd b_2-P[C_3]$
of the worldvolume two-form $b_2$ and its flux sectors enter
\cite{Pasti:1997gx,Bandos:1997ui}.~\footnote{Setting
$H_3=0$ implies $P[G_4]=0$ and a compatible global
trivialization of the pulled-back potential data.}
Known defect and compact-instanton actions provide
further benchmarks \cite{Beccaria:2024gkq,Lin:2026uzp}.

Another direction is to study probes in backgrounds
sourced by branes wrapped on compact cycles with
magnetic charges. Their auxiliary pairs
$(Y_{12},E_4)$ or $(Y_{12},E_6)$ are being developed
in a separate project \cite{Hristov_toappear}.
Applying the probe prescription could give access
to nonperturbative contributions and extended
observables in twisted indices, holographic
renormalization-group flows and black holes.

Finally, equivariant descriptions of bulk and probe
fluctuations
\cite{BenettiGenolini:2026kjy,Kurlyand:2026yke,Kurlyand:2026gxi}
suggest extending our auxiliary framework to characters
and indices. Existing results include M2 fluctuation
characters and localization on brane families
\cite{BenettiGenolini:2026gjn}, as well as localization
and quantization of protected D3 giant-graviton sectors
\cite{Eleftheriou:2025lac,GonzalezLezcano:2026wje}.
Incorporating probe prefactors alongside the classical
exponents would require the appropriate fluctuation
complexes, bundle data and treatment of zero modes.\\

\noindent\textbf{Acknowledgements. }We acknowledge assistance from ChatGPT (OpenAI) with helpful discussions, derivations and editing. K.H.\ is supported in part by the Bulgarian NSF grant KP-06-N68/3. K.H.\ gratefully acknowledges the hospitality and support of the Institute for Advanced Study, where this work was carried out, with funding provided by an anonymous donor. P.V.M.
is supported by the Netherlands Organisation for Scientific Research (NWO) under the
VICI grant VI.C.202.104.

\appendix
\section{Equivariant integration}
\label{app:equivariant}

\emph{Relative localization.}
Let $M$ be a compact oriented $(2n-1)$-manifold without
boundary, carrying an action of a torus $T$.
Assume that $M$ admits a compact oriented filling $W$
extending the torus action and inducing the chosen
orientation on $M$.
Write $j:M=\partial W\hookrightarrow W$ for the
equivariant boundary inclusion.
Let $K$ be the vector field generated by a generic
element of the torus Lie algebra on $W$, and assume
that its restriction to $M$, also denoted $K$, is
nowhere zero.
Its zero set is then $F=W^T$, where $W^T$ denotes
the points fixed by every element of $T$.
The absence of boundary zeros places $F$ entirely
in the interior of $W$. Complex weights are understood by meromorphic continuation
from this compact real torus action.

On invariant polyforms, $\dd_K=\dd-\iota_K$ uses
the vector field generating the torus action on the corresponding space.
Equivariance of $j$ ensures
$j^*\dd_K=\dd_Kj^*$.
A relative cocycle consists of a polyform $\alpha$
on $W$ and a polyform $\beta$ on $M$, satisfying
\begin{equation}
 \dd_K\alpha=0\,,\qquad j^*\alpha=-\dd_K\beta\,.
 \label{eq:relativecocycle}
\end{equation}
Writing $[\cdot]_p$ for ordinary form degree $p$,
its pairing is
\begin{equation}
 \begin{aligned}
J[\alpha,\beta]:=&\int_W[\alpha]_{2n}+\int_M[\beta]_{2n-1}\\
 &=\int_F\frac{\alpha|_F}{e_T(NF)}
 \equiv\int_W^{\rm eq}\alpha.
 \end{aligned}
 \label{eq:relativepair}
\end{equation}
Here $e_T(NF)$ is the equivariant Euler class of the
normal bundle, evaluated at the chosen torus parameters;
integration over $F$ includes all connected components.
A collar extension converts the relative pair into
a closed representative vanishing on the boundary
and unchanged near $F$
\cite[Eqs.~(46)--(48)]{BenettiGenolini:2026cdw}.
Applying ABBV localization
\cite{Atiyah:1984px,BerlineVergne1982Localization}
then includes the boundary term in the fixed-locus
integral \cite{ParadanVergne2007}.
For compact $W$ without boundary, this reduces to
ordinary equivariant integration.

\emph{Relative lifts of odd-dimensional integrals.}
A useful explicit realization of
\eqref{eq:relativecocycle} starts from an invariant
one-form $\chi$ on $M=\partial W$, with
$\dim W=2n$ and $\iota_K\chi=c$ constant and nonzero.
Cartan's identity then gives $\iota_K\dd\chi=0$.
Define
\begin{equation}
 \alpha=c^n\,,\qquad
 \beta=\chi\wedge
       \sum_{j=0}^{n-1}c^{n-1-j}(\dd\chi)^j\,.
 \label{eq:contactrelative}
\end{equation}
The scalar $c$ carries equivariant degree two,
so $\alpha$ and $\beta$ have total degrees $2n$
and $2n-1$. Directly,
$\dd_K\beta=(\dd\chi)^n-c^n=-j^*\alpha$,
since $(\dd\chi)^n$ vanishes on the $(2n-1)$-dimensional
boundary. Thus the odd-dimensional integrand is the
boundary component of a relative class of one higher
total degree. Equation~\eqref{eq:relativepair} becomes
\begin{equation}
 \int_W^{\rm eq}c^n
 =\int_M\chi\wedge(\dd\chi)^{n-1}\,.
 \label{eq:contactpushforward}
\end{equation}

After formally inverting $c$, the same construction gives a
relative lift of the unit. Define the finite series
\begin{equation}
 \begin{aligned}
 \Theta_\chi
 &:=\frac{\beta}{c^n}
   =\frac{\chi}{c-\dd\chi}
   =\sum_{r=0}^{n-1}
      \frac{\chi\wedge(\dd\chi)^r}{c^{r+1}}\,,\\
 \dd_K\Theta_\chi&=-1\,.
 \end{aligned}
 \label{eq:relativeunit}
\end{equation}
Hence $(1,\Theta_\chi)$ is a relative cocycle in
localized equivariant cohomology. More generally, any
even equivariantly closed polyform $P$ on $W$ satisfying $\dd_KP=0$ has the lift
$(P,j^*P\wedge\Theta_\chi)$.
Restriction to a lower-dimensional invariant boundary
cycle simply truncates the series.

Products of these relative pairings
\cite{ParadanVergne2007} have bulk member
$\alpha_1\alpha_2$ and corner member
$\beta_1\beta_2$, with the corresponding face
terms and induced orientations.
Their pairing is the product of the factor pairings.
This is used for both D3 background and probe actions
in Section~\ref{sec:D3}.
For $W_1\times W_2$ with even-dimensional factors, the
corner orientation inherited from the face
$\partial W_1\times W_2$ is minus the product boundary
orientation. Face and corner signs must therefore be
retained together.
These constructions require globally defined invariant
data or compatible gauge-patching terms.

\emph{Equivariant volumes and insertions.}
We use normalized equivariant Chern roots: for a
$2\pi$-periodic angle $\varphi$ and
$K=v\partial_\varphi$, the Euler root is
$\epsilon=v/(2\pi)$.
For example, $\chi=\dd\varphi$ gives
$c=v=2\pi\epsilon$ and $\int_{S^1}\chi=2\pi$,
consistently with \eqref{eq:contactpushforward}.
The physical relative representatives and their
pushforwards must be converted to these conventions together.

Let $X$ be a smooth toric complex space, or a specified
orbifold model, with compact fixed locus. For normalized equivariant divisor classes $x_i$
and formal generating parameters $\lambda^i$, define
\begin{equation}
 \BV_X(\lambda,\epsilon)
=\int_{X^T}\frac{e^{\sum_i\lambda^i x_i}|_{X^T}}
                       {e_T(NX^T)}.
 \label{eq:volumeconvention}
\end{equation}
For toric quotients this can equivalently be computed
by a Jeffrey--Kirwan residue in the chosen chamber
\cite{Jeffrey:1993cun,Martelli:2023oqk,Cassia:2025jkr,Hristov:2026uxs}.
We set $C_X(\epsilon) :=\BV_X(0,\epsilon)$, with
$C_{\BC^d}=1/\prod_i\epsilon_i$. Polynomial insertions of equivariant Chern roots are generated by
\begin{equation}
 \int_X^{\mathbb V}P(x,\epsilon):=\left.P(\partial_\lambda,\epsilon)\BV_X(\lambda,\epsilon)\right|_{\lambda=0}\,.
 \label{eq:volumeregulator}
\end{equation}
For noncompact $X$, this defines a meromorphic
pushforward, finite at generic weights.
Insertions may have meromorphic coefficients in
the equivariant parameters. Derivatives are evaluated
before restriction to a smaller parameter locus. At $\lambda=0$, a complex $d$-dimensional
cycle has $C_D(-\epsilon)=(-1)^dC_D(\epsilon)$.
Characteristic classes are inserted as polynomials in
the $x_i$; for example, the toric Euler sequence gives
$p_1(TX)=\sum_i x_i^2$ and
$p_2(TX)=\sum_{i<j}x_i^2x_j^2$.

Let $\iota_D:D\hookrightarrow X$ be an invariant
oriented inclusion of real codimension $2s$.
Its equivariant cycle class $[D]_T$ is the equivariant
Poincar\'e dual, obtained by pushing forward the unit
class on $D$ using the Thom class of its normal bundle
\cite{Fujisawa2018}.
For a regular intersection of invariant divisors,
$D=D_{i_1}\cap\cdots\cap D_{i_s}$, with its induced
intersection orientation, it is
\begin{equation}
 [D]_T:=(\iota_D)_*1
 =\prod_{a=1}^{s}x_{i_a}\,.
 \label{eq:equivariantcycleclass}
\end{equation}
Here $(\iota_D)_*$ is the Gysin pushforward from $D$
to $X$, while $\iota_D^*$ is restriction from $X$ to $D$.
Thus $\iota_D^*1=1$, whereas $[D]_T$ has equivariant
degree $2s$ and encodes the normal directions to $D$.
For a smooth inclusion, its restriction to $D$ is
the normal Euler class, $\iota_D^*[D]_T=e_T(ND)$.
Intersection and orbifold multiplicities are included
in the cycle classes and pushforwards.

The cycle volume is therefore computed by inserting
$[D]_T$ in the ambient equivariant volume:
\begin{equation}
 \begin{aligned}
 C_D(\epsilon)=\int_D^{\mathbb V}1
 &=\int_X^{\mathbb V}[D]_T\\
 &=\left.\prod_{a=1}^{s}\partial_{\lambda^{i_a}}
 \BV_X(\lambda,\epsilon)\right|_{\lambda=0}.
 \end{aligned}
 \label{eq:cyclevolume}
\end{equation}
A polynomial insertion $P(x,\epsilon)$ on the same cycle
obeys the projection formula:
\begin{equation}
 \begin{aligned}
 \int_D^{\mathbb V}\iota_D^*P
 &=\int_X^{\mathbb V}P[D]_T\\
 &=\left.
 P(\partial_\lambda,\epsilon)
 \prod_{a=1}^{s}\partial_{\lambda^{i_a}}
 \BV_X(\lambda,\epsilon)
 \right|_{\lambda=0}.
 \end{aligned}
 \label{eq:projectionformula}
\end{equation}
Thus $P=1$ gives the cycle volume, while $P=x_j$ adds
one derivative $\partial_{\lambda^j}$.
For a more general invariant cycle, the product of divisor
classes is replaced by its equivariant cycle class.\\

\bibliography{refs}
\end{document}